\pdfoutput=1
\documentclass[12pt,letterpaper]{article}
\usepackage{jheppub}

\def\l{\langle}
\def\r{\rangle}
\def\nl{\nonumber\\}
\def\o{\sigma}

\DeclareMathOperator{\Conf}{Conf}
\DeclareMathOperator{\Gr}{Gr}
\DeclareMathOperator{\Li}{Li}
\DeclareMathOperator{\li}{li}

\title{The Differential of All Two-Loop MHV Amplitudes in $\mathcal{N}=4$ Yang-Mills Theory}

\author[a]{John Golden,}
\author[a,b]{Marcus Spradlin}

\affiliation[a]{Department of Physics, Brown University, Box 1843,\\
Providence, RI 02912-1843, USA}
\affiliation[b]{Theory Division, Physics Department, CERN,\\
1211 Geneva 23, Switzerland}

\abstract{
We present an explicit analytic calculation of the differential of the planar $n$-particle, two-loop MHV scattering amplitude in $\mathcal{N}=4$ super Yang-Mills theory. The result is expressed only in terms of the polylogarithm functions $\Li_k(-x)$, for $k=1,2,3$, with arguments $x$ belonging to the special class of dual conformal cross-ratios known as cluster $\mathcal{X}$-coordinates. The surprising fact that these amplitudes may be expressed in this way provides a striking example of the manner in which the cluster structure on the kinematic configuration space underlies the structure of amplitudes in SYM theory.\vspace{6cm}
}
\notoc
\begin{document}
\maketitle

\pagebreak

\section{Introduction}

Insights gleaned from difficult calculations have driven much of the recent progress in unlocking the structure of
$\mathcal{N}=4$ super Yang-Mills (SYM) theory.
The synergy between technical and conceptual
advances frequently enables the undertaking of previously
impossible calculations; the often surprising
results of which in turn fuel a deeper understanding of the theory. 
Indeed, in SYM theory the search for such insights, rather than the need
for busywork, is precisely the motivation for carrying out arduous
new calculations. 

In this note we complete the analytic calculation of the differential
$dR_n^{(2)}$ of the planar two-loop $n$-particle MHV amplitudes
(or, more properly, the `remainder functions'---with
infrared divergences
subtracted off in a standard way~\cite{Anastasiou:2003kj,Bern:2005iz}).
Reasonably efficient numerical techniques for evaluating
$R_n^{(2)}$ have been available for some time~\cite{Anastasiou:2009kna},
and analytic formulas are known in two-dimensional
kinematics~\cite{Heslop:2010kq}
but a breakthrough towards unlocking the general analytic structure of these amplitudes
was made by Caron-Huot in~\cite{CaronHuot:2011ky}.
By considering the dual superconformal symmetry of a certain
generalization of scattering amplitudes depending on twice the usual
number of Grassmann variables, he was able to
express $dR_n^{(2)}$ in terms of a certain combination of one-fold integrals.
Here we complete the evaluation of these integrals and present
analytic expressions for $dR_n^{(2)}$. We find that they
can be expressed completely in terms of the functions $\Li_k(-x)$,
for $k=1,2,3$ with arguments $x$ always belonging to the set of
cluster $\mathcal{X}$-coordinates on the kinematic
configuration space $\Conf_n(\mathbb{P}^3)$.
This provides strong support for the suggestion of~\cite{Golden:2013xva}
that the cluster structure of this space underlies the structure of
amplitudes in SYM theory.

While this computation may seem modest in scope, we feel that it is a 
useful example in which to showcase the power of the two
most recent additions to the
amplitudeologist's toolkit:
positivity and cluster $\mathcal{X}$-coordinates.
In particular,
the positive domain,
a subset of the
kinematic domain $\Conf_n(\mathbb{P}^3)$, is evidently the natural
habitat for amplitudes in SYM theory, and
cluster $\mathcal{X}$-coordinates provide a natural set of arguments
for these amplitudes to `depend on'.
It is our hope that other analytic formulae in SYM theory can be unlocked using these or similar approaches. 
In particular, it would be very interesting to see if the monodromies of
$R_n^{(2)}$, which were computed in integral form in~\cite{Gaiotto:2011dt}
(see also~\cite{Sever:2011pc}),
can be similarly expressed only in terms of $\Li_k(-x)$.

Following ample motivation (but minimal review), we present our analytic expression for $dR_n^{(2)}$. For additional introductory material, we refer the reader to~\cite{Golden:2013xva} and the references therein. 

\section{Motivation and Review}

In~\cite{CaronHuot:2011ky}, Caron-Huot expressed the differential of the planar two-loop $n$-particle MHV amplitude in terms of momentum twistors \cite{Hodges:2009hk}  as
\begin{equation}
\label{eq:simon}
 d R^{(2)}_n = \sum_{i,j} C_{i,j}~d\log \langle i{-}1\,i\,i{+}1\,j\rangle
 \end{equation}
 and presented a means of calculating the coefficient functions $C_{i,j}$.
Already the fact that the differential may be expressed in the
form of eq.~(\ref{eq:simon}) is a nontrivial all-loop-order prediction
of~\cite{CaronHuot:2011ky}.

Thanks to the full dihedral symmetry of the amplitude 
in the particle labels, it is sufficient to focus on
$C_{2,i}$.
This was given as a
sum of four components $C_{2,i}^{(a)}$, for $a=1,2,3,4$,
two of which were represented analytically as
\begin{equation}\label{c1}
 C_{2,i}^{(1)}= \log u \times \sum_{j=2}^{i{-}1} \sum_{k=i}^{n{+}1}
 \left[\Li_2(1-u_{j,k,k{-}1,j{+}1}) +
  \log \frac{x_{j,k}^2}{x_{j{+}1,k}^2} \log \frac{x_{j,k}^2}{x_{j,k{-}1}^2}\right]
\end{equation}
where for our purposes it is sufficient to take
$x_{i,j}^2 = \langle i\,i{+1}\,j\,j{+}1\rangle$, and
\begin{equation}\label{c4}
  C_{2,i}^{(4)} =  -2\Li_3(1-\frac{1}{u}) - \Li_2(1-\frac1u)\log u - \frac16 \log^3u + \frac{\pi^2}{6} \log u,
\end{equation}
where
\begin{equation}
 u_{i,j,k,l} = \frac{\l i \,i{+}1\, j\,j{+}1\r \l k\,k{+}1\, l\,l{+}1\r}{\l  i \,i{+}1\,k\,k{+}1\r \l j\,j{+}1 \, l\,l{+}1\r}
\end{equation}
 and $u \equiv u_{2,i{-}1,i,1}$. $C^{(2)}_{2,i}$ has the integral representation
\begin{eqnarray}\label{c2}
 C^{(2)}_{2,i}&=&
 \phantom{+}\sum_{j=4}^{i{-}2} \int _1^2 d_X\!\log\frac{\l Xi{-}1i\r}{\l Xij\r} \Big{(}\Li_2(1-u_{X,i{-}1,i,j}) - \Li_2(1-u_{X,i{-}1,i,j{-}1}) + \Li_2(1-u_{i,j{-}1,j,i{-}1}) \nl &&
\hspace{2cm}  + \Li_2(1-u_{X,j,j{-}1,i{-}1}) -\Li_2(1-u_{X,j,j{-}1,i}) + \log u_{X,i{-}1,i,j}\log u_{X,j,j{-}1,i{-}1}\Big{)}\nl
 \nl && + \sum_{j=4}^{i{-}2} \int_1^2 d_X\!\log\frac{\l Xi{-}1i\r}{\l Xii{+}1\r} \Big{(} \Li_2(1-u_{i{-}1,1,X,j}) + \Li_2(1-u_{i,j,j{-}1,1}) + \Li_2(1-u_{1,j,j{-}1,X})
 \nl && \hspace{2cm} -\Li_2(1-u_{i{-}1,1,X,j{-}1}) - \Li_2(1-u_{i,j,j{-}1,X}) + \log u_{i{-}1,1,X,j} \log u_{i,j,j{-}1,1}\Big{)}
  \nl \nl &&
 + \int_1^2 d_X\!\log\frac{\l Xi{-}1i\r}{\l X34\r} \left( \Li_2(1-u_{2,i,i{-}1,3})-\Li_2(1-u_{X,i,i{-}1,3})\right)
\end{eqnarray}
and $C^{(3)}_{2,i}$ may be determined from $C^{(2)}_{2,i}$
by imposing dihedral
symmetry on eq.~(\ref{eq:simon}), which fixes
\begin{equation}
C^{(3)}_{2,i}(1,2,\ldots,n) = C^{(2)}_{2,4+n-i}(3,2,\ldots,4).
\end{equation}

While eqs.~(\ref{c1}) and~(\ref{c4}) along with the integral representation in eq.~(\ref{c2}) were sufficient to compute the symbol of $d R^{(2)}_n$ in~\cite{CaronHuot:2011ky}, any attempt at an analytic expression for $d R^{(2)}_n$ would have encountered two major obstacles:  

\subsection*{Choosing a `Good' Kinematic Domain}

First of all, $n$-particle
scattering amplitudes in SYM theory are multi-valued functions
on the
$3(n-5)$-dimensional kinematic configuration space $\Conf_n(\mathbb{P}^3)$
(we refer the reader to~\cite{Golden:2013xva} for a review of this notation),
and in computing scattering amplitudes it is often very difficult
to get all of the branch cuts in the right place. In particular, Mathematica's {\tt Integrate[]} function 
has no \emph{a priori} understanding of $\Conf_n(\mathbb{P}^3)$ and cannot be expected to easily 
return an analytic formula valid on the whole of this domain---at best
it can produce numerical results at specifically inputted kinematic
points, or an analytic formula valid on some subdomain of its own choosing.
On physical grounds, amplitudes should be real-valued and
singularity-free throughout the Euclidean domain,
a subset of $\Conf_n(\mathbb{P}^3)$.
Even for what is in some sense the simplest
non-trivial multi-loop amplitude, $R_6^{(2)}$, finding a formula
with these properties was by far the hardest part of~\cite{Goncharov:2010jf}.
For $n>6$, where the kinematic configuration space is far more
complicated (due to Gram determinant constraints amongst dual conformal
cross-ratios),
finding explicit formulas valid throughout even just the Euclidean domain
seems like a daunting challenge.
Instead of giving up all hope, one would be content to find
expressions valid even in some subset of the Euclidean domain.
But, which subset should one look at?
What principle, either mathematical or physical, could
make any one subset more worthy of attention than another?

\subsection*{Choosing `Good' $\Li_k$ Arguments}

The second complication has to do with the class of iterated integrals
of the type which define generalized polylogarithm functions and which
appear in the result of~\cite{CaronHuot:2011ky}.
Such functions can be partially characterized by their symbols, but integrating a function of this type can generate
a function whose symbol contains entries that are, in general,
arbitrary alebgraic functions of the entries in the symbol of the
original function.  Already Caron-Huot's result for the symbol
of $R_n^{(2)}$ contains rather non-trivial algebraic functions on
$\Conf_n(\mathbb{P}^3)$, and one might have worried that even more
complicated functions could appear after integration. Yet, true believers in
SYM theory know well that it only
ever produces very special polylogarithmic functions, not `general' ones. One would therefore
 have every reason to expect only relatively straightforward expressions to
actually appear in amplitudes.
Unfortunately, unless one knows in advance what to look out for,
the plethora of identities amongst polylogarithm functions
makes it difficult to identify any particular set of arguments as
preferred over any other set.

\section{Tools and Techniques}

Fortunately, related recent advances have shown how to overcome both obstacles
at the same time.

\subsection*{Cluster $\mathcal{X}$-coordinates}

We begin by addressing the question of $\Li_k$ arguments by noting that
in~\cite{Golden:2013xva} it was shown via examples at $n=6,7$, and
suggested more generally, that (in a sense made precise in that paper)
two-loop MHV amplitudes only `depend on' certain very special variables. 
These variables are called cluster $\mathcal{X}$-coordinates on the kinematic
configuration space $\Conf_n(\mathbb{P}^3)$, which may be realized
as the Grassmannian quotient $\Gr(4,n)/(\mathbb{C}^*)^{n-1}$.
For the purpose of this paper, the notion of `depending on' only certain
variables can be stated very clearly: in support of~\cite{Golden:2013xva},
we find that the differential $d R_n^{(2)}$ of all two-loop MHV
amplitudes can be expressed only in terms of the functions $\Li_k(-x)$,
for $k=1,2,3$ with arguments $x$ always belonging to the set of
cluster $\mathcal{X}$-coordinates on $\Conf_n(\mathbb{P}^3)$.

In fact, we find that (in line with the expectation expressed in the
previous section) only relatively
tame $\mathcal{X}$-coordinates appear in $dR^{(2)}_n$, for any $n$.
The most complicated of these is the
cross-ratio
\begin{equation}
\frac{\langle 12i{-}1i\rangle  \langle 123j\rangle
\langle 2j{-}1jj{+}1\rangle  \langle j{-}1ji{-}1i\rangle }
{\langle 12j{-}1j\rangle  \langle 2ji{-}1i\rangle
\langle i{-}1i(123){\cap}(j{-}1jj{+}1)\rangle }
\end{equation}
which depends on the eight points $\{1,2,3,j-1,j,j+1,i-1,i\}$. Here we use the notation
\begin{equation} 
  \langle a b (c d e) \cap (f g h)\rangle \equiv \langle a c d e\rangle \langle b f g h\rangle - \langle b c d e\rangle \langle a f g h\rangle.
\end{equation} 
The fact that the complexity
of two-loop MHV amplitudes
stabilizes (in this sense) already at $n=8$ is manifest in the
result of~\cite{CaronHuot:2011ky}.
This implies that a full understanding of
the cluster structure on $\Conf_8(\mathbb{P}^3)$ should be sufficient to
understand the structure of
all such amplitudes.

\subsection*{The Positive Domain}

We now turn to the question of the kinematical domain, where the sole dependence on cluster $\mathcal{X}$-coordinates plays a fortuitous role.
A salient feature of cluster $\mathcal{X}$-coordinates is that they are
positive-valued everywhere in the positive domain, which is the
subset of the Euclidean domain defined by imposing that
$\l a b c d \r > 0$ whenever $a < b < c < d$.
With the differential $dR_n^{(2)}$ expressed completely in terms of the
functions $\Li_k(-x)$, it therefore becomes manifestly real-valued and
singularity free throughout the positive domain.  (Moreover, we believe
that $R_n^{(2)}$ itself is also positive-valued throughout the positive
domain, but that is another matter.)
The importance of positivity for scattering amplitudes was first
emphasized in~\cite{ArkaniHamed:2012nw}, though in a somewhat different
context.
Their Grassmannian
for the integrand
involves both `external' kinematic data as well as `internal' loop
integration variables,
but it is now clear that positivity in the `external' data
alone plays a similarly important role for the fully integrated
amplitudes which we study here.

Before proceeding let us add one crucial disclaimer.  As reviewed
in~\cite{Golden:2013xva}, for $n>7$ the cluster algebra associated
to $\Conf_n(\mathbb{P}^3)$ has an infinite number of cluster
$\mathcal{X}$-coordinates (although of course only a finite number
actually
appear in $dR^{(2)}_n$).
Moreover there is no known general classification of these coordinates,
and given any cross-ratio there does not exist a
general algorithm for determining whether or not it actually
is a cluster $\mathcal{X}$-coordinate.
Therefore, for $n>7$ we use the empirical criterion
discussed in section 6.6 of~\cite{Golden:2013xva}:
we say that a cross-ratio $x$ is a cluster $\mathcal{X}$-coordinate
if $1 + x$ factors into a product of four-brackets as a consequence
of Pl\"ucker relations and if $x>0$ everywhere
inside the positive domain.

\vskip .3cm

\noindent
Given the two advances we have
reviewed---the understanding that the natural domain
on which to study multi-loop amplitudes in SYM theory is the positive
domain, and that the natural set of arguments appearing inside the $\Li_k$
functions for two-loop MHV amplitudes are cluster
$\mathcal{X}$-coordinates---analytically integrating eq.~(\ref{c2}) transforms from being merely possible to
being inevitable.

\section{Results}
In this section we present the $\Li_3$ and $\Li_2 \times \Li_1$ contributions to $dR^{(2)}_n$. There are, in addition, terms of the form $\Li_1^3$ and $\pi^2 \Li_1$ which are too numerous to efficiently display here.  For this reason we attach to this submission a Mathematica notebook which contains the necessary expressions and which can construct the full analytic formula for $dR^{(2)}_n$ for any given $n$.

While the original integral representation of $C_{i,j}$ required the general object $u_{i,j,k,l}$, the integrate{\it d} result can be written in terms of the slightly smaller class of cross-ratios defined by $u_{i,j,k} = u_{i,j,k,i{-}1}$. The cluster $\mathcal{X}$-coordinates related to the three-index $u$'s (in the same sense that the $v_i$ and the $u_i$ used in~\cite{Golden:2013xva} for $n=6$ are related to each other) are
\begin{equation}
v_{i,j,k} = \frac{1}{u_{i,j,k}}-1 = -\frac{\l i (i{-}1\,i{+}1)(j\,j{+}1)(k\,k{+}1) \r}{\l i{-}1\,i\,k\,k{+}1\r \l i\,i{+}1\,j\,j{+}1\r}.
\end{equation}
Here we use the notation
\begin{equation} 
  \langle a (b c) (d e) (f g) \rangle \equiv \langle a b d e\rangle \langle a c f g\rangle - \langle a b f g\rangle \langle a c d e\rangle.
\end{equation} 
For any $n$, $v_{i,j,k}$ is a cluster $\mathcal{X}$-coordinate as long as $i<j<k$ (mod $n$). We also find it useful to define another type of ratio,
\begin{equation}
w_{i,j,k} = \frac{v_{i,k{-}1,k} u_{i,k{-}1,k}}{v_{j,k{-}1,k}} 
\end{equation}
which is a cluster $\mathcal{X}$-coordinate when $i=j\pm 1$. Because we are interested only in objects of the form $\Li(-x)$, where $x$ is a cluster $\mathcal{X}$-coordinate, we will make one more definition
\begin{equation}
\li_k(x) = \Li_k(-x)
\end{equation}
purely in the interest of cleaning up the notation. 

We begin with the essentially semantic task of converting $C^{(1)}$ and $C^{(4)}$ into $\mathcal{X}$-coordinate form. Expressing $C_{2,i}^{(4)}$ in the desired form is trivial:
\begin{equation}\label{c4full}
 C_{2,i}^{(4)} =  -2\li_3(v_{2,i{-}1,i}) - \li_2(v_{2,i{-}1,i})\li_1 (v_{2,i{-}1,i}) +\mathcal{O}(\li_1^3).
  \end{equation}
 $C^{(1)}$ is slightly more difficult, since it involves non-conformal objects such as $\log \frac{x_{j,k}^2}{x_{j,k{-}1}^2}$. Of course, the expression as a whole is conformally invariant, but it takes some combinatoric effort to recast things as an explicit sum over conformal cross-ratios. This is only a problem for the $\mathcal{O}(\Li_1^3)$ ratios, our results for this are in the attached Mathematica notebook. For $\Li_2\times \Li_1$, we find that the ratios appearing are in fact already in cluster $\mathcal{X}$-coordinate form, giving
\begin{eqnarray}\label{c1full}
 C_{2,i}^{(1)} =  -\li_1(v_{2,i{-}1,i}) \Bigg{(}&&\sum_{j=2}^{i-2}\sum_{k=i}^n \li_2(v_{j{+}1,k,k{+}1})\nl&&-\sum_{k=i+2}^n \li_2(v_{k,2,i{-}1})-\sum_{j=4}^{i-2}\li_2(v_{j,i,1}) \Bigg{) +\mathcal{O}(\li_1^3). }
\end{eqnarray}
This sum looks slightly different than eq.~(\ref{c1}) because we find it useful to make explicit the behavior of $1-u_{j,k,k{-}1,j{+}1}$ at various upper and lower limits of the $j,k$ summation indices.
The double-sum of eq.~(\ref{c1full}) contains some boundary terms which
diverge---specifically, when $j,k$ equal $2,n$ or $i-2,i$.
However, these divergences cancel when the $\mathcal{O}(\Li_1^3)$ terms are
added.  The attached Mathematica file identifies and discards these canceling
divergent terms.

We now turn to the main focus of this note: the evaluation of the remaining integrals in $C^{(2)}_{2,i}$.  For fixed $i,j$, the terms on the first four lines of eq.~(\ref{c2}) altogether depend on at most nine distinct momentum twistors. We can therefore focus our efforts entirely on evaluating the integrals for the case $n=9, i=8, j=5$, and then replace $\{ 4,5,6,7,8,9\} \to \{j-1,j,j+1,i-1,i,i+1\}$ to recover the summand for arbitrary $i,j$. One may immediately worry about divergences at boundary terms in the sum, such as $j=i-2$; while these cases do create individual terms that are divergent, these divergences cancel out in the full function. 

In order to present our result for $C^{(2)}_{2,i}$ in a succinct fashion, let us define the following permutation operators acting on the particle labels:
\begin{eqnarray}\label{symmetries}
\o_a&&= Z_{a{-}1} \leftrightarrow Z_{a{+}1}\\
\o_{a,b}&&=1-\o_a-\o_b+\o_a\o_b\\
\o_{a,b,c} &&=1 - \o_a - \o_b - \o_c + \o_a\o_b+\o_a\o_c +\o_b\o_c - \o_a\o_b\o_c
\end{eqnarray}
as well as the four cross-ratios
\begin{eqnarray}\label{ratios}
R_1&=&\frac{\langle 12j{-}1j\rangle  \langle 2ji{-}1i\rangle  \langle j{-}1jj{+}1i\rangle }{\langle 2j{-}1jj{+}1\rangle  \langle 12ji\rangle  \langle j{-}1ji{-}1i\rangle },\nl
R_2&=&\frac{\langle 2i{-}1ii{+}1\rangle  \langle 12ij\rangle  \langle i{-}1ij{-}1j\rangle }{\langle 12i{-}1i\rangle  \langle 2ij{-}1j\rangle  \langle  i{-}1ii{+}1j\rangle },\nl
R_3&=&\frac{\langle 123j\rangle\langle 12i{-}1i\rangle   \langle 2j{-}1ji\rangle }{\langle 123i\rangle  \langle 12j{-}1j\rangle  \langle 2i{-}1ij\rangle },
\nl
R_4&=&\frac{\langle 12i{-}1i\rangle  \langle 123j\rangle  \langle 2j{-}1jj{+}1\rangle  \langle j{-}1ji{-}1i\rangle }{\langle 12j{-}1j\rangle  \langle 2ji{-}1i\rangle  \langle i{-}1i(123){\cap}(j{-}1jj{+}1)\rangle }.
\end{eqnarray}
These ratios have the following $1+R_i$ factorizations:
\begin{eqnarray}
1+R_1&=&\frac{\langle 2j{-}1ji\rangle  \langle j(j{-}1j{+}1)(12)(i{-}1i)\rangle}{\langle 2j{-}1jj{+}1\rangle  \langle 12ji\rangle  \langle j{-}1ji{-}1i\rangle },\nl
1+R_2&=&\frac{\langle 2ji{-}1i\rangle  \langle i(i{-}1i{+}1)(12)(j{-}1j)\rangle}{\langle 12i{-}1i\rangle  \langle 2ij{-}1j\rangle  \langle i{-}1ii{+}1j\rangle },\nl
1+R_3&=&-\frac{\langle 12ji\rangle  \langle 2(13)(j{-}1j)(i{-}1i)\rangle}{\langle 123i\rangle  \langle 12j{-}1j\rangle  \langle 2ji{-}1i\rangle },
\nl
1+R_4&=&-\frac{\langle 2(13)(j{-}1j)(i{-}1i)\rangle \langle j(j{-}1j{+}1)(12)(i{-}1i)}{\langle 12j{-}1j\rangle  \langle 2ji{-}1i\rangle  \langle i{-}1i(123){\cap}(j{-}1jj{+}1)\rangle },
\end{eqnarray}
None of the $R_i$ depend on more than eight points, and we have checked that they are all $\mathcal{X}$-coordinates of the $\text{Gr}(4,8)$ cluster algebra (for more information on generating cluster $\mathcal{X}$-coordinates via mutation, see \cite{Golden:2013xva}).

Note that there is additional symmetry present in in the arguments of $R_1$ and $R_2$, which are related via $R \to 1/R$ combined with $ i\leftrightarrow j$. 

Given these definitions, we find that $C^{(2)}_{2,i}$ has the following relatively simple $\Li_3$ contribution:
\begin{eqnarray}\label{c2li3}
&&\li_3(v_{2,i{-}1,i})-\li_3\left(w_{3,2,i}\right)+\li_3\left(w_{2,3,i}\right)\nl
&&\qquad
\qquad
\qquad
\qquad
-\sum_{j=4}^{i-2} \o_{2,i,j} \Big{(}\li_3\left(R_1\right)+\li_3\left(R_2\right)+\li_3\left(R_3\right)-\li_3\left(R_4\right)\Big{)}.
\end{eqnarray}
The three symmetries lead to 32 distinct $\Li_3$'s in each term
in the $j$ sum. This presentation has discarded all terms which cancel telescopically in the sum. One can alternatively incorporate the non-summand terms into the summand by including some telescopic cancellations, thus increasing the total number of $\Li_3$ terms (still all with cluster $\mathcal{X}$-coordinates as arguments) in the summand to 48. 

While there are no remaining telescopic cancellations in eq (\ref{c2li3}), the full 32 $\Li_3$ terms do not appear for all values of $i$ and $j$\footnote{The exact counting is: for $n>6$ the total number of $\Li_3$ terms in $C^{(2)}_{2,i}$ is given by $32 i - 169$  for $5<i<n$ and $26 n - 141$ for $i=n$. For $n=6$, $C^{(2)}_{2,6}$ has 11 $\Li_3$ terms. $C^{(2)}_{2,i}=0$ for $i\le5$.}. For example, six of the $\Li_3$ terms go to zero at the upper limit of the sum, $j=i-2$. 

It is interesting to note that the evaluation of the integrals in eq.~(\ref{c2}), for the general case $n=9, i=8, j=5$, necessarily produces $\Li_3$'s with \emph{non}-cluster arguments. However, these terms cancel telescopically in the $j$ sum and are zero at the $j=4$ and $j=i-2$ boundaries. 

Next we turn to the $\Li_2\times \Li_1$ contribution to $C^{(2)}_{2,i}$, which we represent here as
\begin{eqnarray}\label{c2li2}
&&\hspace{-1.5cm}\li_1\left(v_{2,i-1,i}\right) \left(\li_2(v_{i,1,3})-\li_2\left(v_{3,i-1,i}\right)\right)-\left(\li_1\left(v_{2,i-1,i}\right)+\li_1\left(v_{3,i-1,i}\right)\right) \li_2\left(w_{2,3,i}\right)\nl
&&\hspace{-1.5cm}+\sum_{j=4}^{i-2}\sigma_{2,j}\Bigg{(}
  \li_1\left(v_{i,2,j-1}\right) \left(\li_2\left(R_2\right) - \li_2\left(R_3\right)\right) + 
   \li_1\left(v_{i,1,2}\right) \sigma_i \li_2\left(R_2\right) \nl
   &&\hspace{-1cm}+ \frac{1}{2}\left(\li_1\left(v_{j,i-1,1}\right) + \left(\frac{1}{2} \li_1\left(v_{j,1,2}\right)-\li_1\left(v_{2,j-1,i-1}\right)
         \right) (1-\sigma_i)- 
      \li_1\left(v_{j,i,1}\right) \sigma_i\right)(\li_2\left(R_1\right) - \li_2\left(R_3\right) + \li_2\left(R_4\right))\Bigg{)} \nl \nl &&\hspace{-1cm}+ 
 \li_1\left(v_{i,1,2}\right) (1-\sigma_j)((1-\sigma_i)\sigma_2\li_2\left(R_2\right)-(\sigma_2+\sigma_i) \li_2\left(R_3\right) + \sigma_i\li_2\left(R_4\right)).
\end{eqnarray}

Note that $\sigma_a v_{a,b,c}$ is not a cluster $\mathcal{X}$-coordinate, so some of the terms in eq. (\ref{c2li2}) aren't obviously in the form $\li_k(x)$. However, this obstruction is easily removed by the relationship $\sigma_a \li_1(v_{a,b,c}) = -\li_1(v_{a,b,c})$. 

We have checked our full result for $C^{(2)}_{2,i}$ against numerical integrations of (\ref{c2}) for several hundred random kinematic points in the positive Grassmannian for $n\le 12$. 

It is important to emphasize that our particular expression for eqs. (\ref{c2li2}) is far from being unique. Some additional symmetry can be introduced by adding terms that telescopically cancel, as was the case with the $\Li_3$ sum. Furthermore, there exist numerous non-trivial identities amongst the $\Li_2$ terms in eq. (\ref{c2li2}), introducing additional redundancy. This particular representation was chosen simply because it was the most typographically concise presentation we could deduce. 

We conclude this note with a brief discussion of parity (see Appendix A of~\cite{Golden:2013xva} for a thorough introduction). The parity invariance of $dR^{(2)}_n$ follows from the fact that $*C_{i,j}=C_{j,i}$ and $\sum_i C_{i,j} = 0$. These properties can be made manifest at the level of the symbol, as was done in~\cite{CaronHuot:2011ky}, but our integrated results sacrifice explicit parity invariance for brevity. Of course, parity invariance in our functional representation can be confirmed through the application of (numerous) polylogarithm identities. 

\acknowledgments

We are grateful to A.~B.~Goncharov, C.~Vergu and A.~Volovich for
closely related
collaboration, and to N.~Arkani-Hamed for proselytizing us to positivity.
This work was supported by the US Department of Energy under contract
DE-FG02-13ER42023.

\end{document}